\documentclass[%
 reprint,
 amsmath,amssymb,
 aps,
]{revtex4-2}

\usepackage{graphicx}
\usepackage{dcolumn}
\usepackage{bm}

\begin{document}

\preprint{APS/123-QED}

\title{Orbital angular momentum enhanced laser absorption and neutron generation}

\author{N. Peskosky}

\author{N. Ernst}

\author{M. Burger}

\author{J. Murphy}

\author{J. Nees}

\author{I. Jovanovic}

\author{A.G.R. Thomas}

\author{K. Krushelnick}%
\email{kmkr@umich.edu}

\affiliation{%
 Gérard Mourou Center for Ultrafast Optical Science, University of Michigan, Ann Arbor, 48109, USA }%

\date{\today}

\begin{abstract}
We experimentally demonstrate enhanced absorption of near relativistic optical vortex beams in D$_2$O plasmas to generate a record fast-neutron yield of 1.45 $\times 10^{6}$ n/s/sr. Beams with a topological charge of 5 were shown to deliver up to a 3.3 times enhancement of fast-neutron yield over a Gaussian focused beam of the same energy but having two orders of magnitude higher intensity. This result was achieved with laser energies of 16 mJ and a pulse duration of 67 fs. The Orbital Angular Momentum (OAM) beam-target interactions in our experiment were also investigated through Particle-in-Cell (PIC) simulations. Electron density rippling resulting in enhanced plasma wave excitation on the critical surface and significantly enhanced resonance absorption is observed. 

\end{abstract}

\maketitle

Vortices are most often associated with the natural motion of fluid systems, examples include hurricanes, tornadoes and whirlpools. Like fluids, electromagnetic waves can exhibit vortex motion and carry orbital angular momentum (OAM)~\cite{Longman22}. The concept of an optical vortex was first described by Nye and Berry~\cite{Nye}, but it was the seminal work by Allen et al.~\cite{Allen92} that showed paraxial Laguerre-Gaussian (LG) modes carry a well defined OAM~\cite{Gruneisen08}. LG modes mathematically characterize circularly symmetric laser beam profiles through cylindrical paraxial solutions of the Helmholtz equation. One of the most unique characteristics of OAM beams is that the twisting helical phase structure creates an on-axis phase singularity. This singularity manifests as an intensity null which gives such beams a ‘doughnut-shaped’ far-field intensity pattern~\cite{Sueda04}. Recently, there has been considerable interest in extending the application of optical vortices to the realm of ultrashort high-intensity lasers. This paper is primarily focused on exploring plasma dynamics resulting from an inward-facing ponderomotive component associated with the on-axis phase singularity present in OAM beams~\cite{Longman22,Dong22}.

In this work, we investigate the creation of high repetition tightly focused (f/1.5) LG$_{01}$ and LG$_{05}$ ultrashort laser beams carrying orbital angular momentum with near relativistic intensity. We experimentally demonstrate deuteron heating using OAM laser beams incident on a microscale free-flowing D$_2$O liquid jet. The high-order LG modes propagating at grazing incidence with respect to overdense targets result in an order of magnitude enhancement in fast neutron generation over previously reported normal incidence interactions using relativistic Gaussian beams of the same energy but two orders of magnitude lower intensity. Through the use of PIC computational methods, we corroborate these experimental findings and confirm evidence of LG beam self-focusing due to nonlinear plasma effects. Following beam collapse, intense polar filamentation is shown to result in significantly enhanced laser absorption into plasma waves and thus into fast particles and subsequently enhanced D(d,n)$^3$He fusion yield as the energy of the electrons is transferred to the ions in the target. Polarization effects on the OAM ion acceleration dynamics are also investigated.
By convention, vortex light beams with finite angular momentum are characterized by an angular-dependent phase $\Phi$ = $l\phi$, a helical wavefront structure having an intensity null on the propagation axis. $l$ is the topological charge that corresponds to the OAM carried by each photon and $\phi$ is the azimuthal angle. Mathematically, these beams are described by Laguerre-Gaussian (LG) modes, and physical implementation usually requires some method of mode conversion from Hermite-Gaussian to LG modes~\cite{Ju18}.
Established methods for generating vortex light beams include beam structuring performed by q-plates~\cite{Marrucci96,Qu17}, digital spatial light modulators (SLMs)~\cite{Gruneisen08,Andersen19}, spiral phase plates (SPP) ~\cite{Sueda04,Oemrawsingh04}, and most recently, dispersionless spiral phase mirrors (SPM) ~\cite{Longman20}. For high intensity applications, such as the experiments performed during this work, spiral phase optics (SPPs and SPMs) provide the primary means of HG to LG conversion. This is due to their high optical damage threshold and ability to mode convert optical pulses having substantial compressed spectral bandwidth, typical of femtosecond laser systems, consistent with pulse durations $<$ 100 fs~\cite{Longman22}, without significant accumulation of second and third order phase.

\begin{figure}[htbp]
  \centering
  \includegraphics[width=\columnwidth]{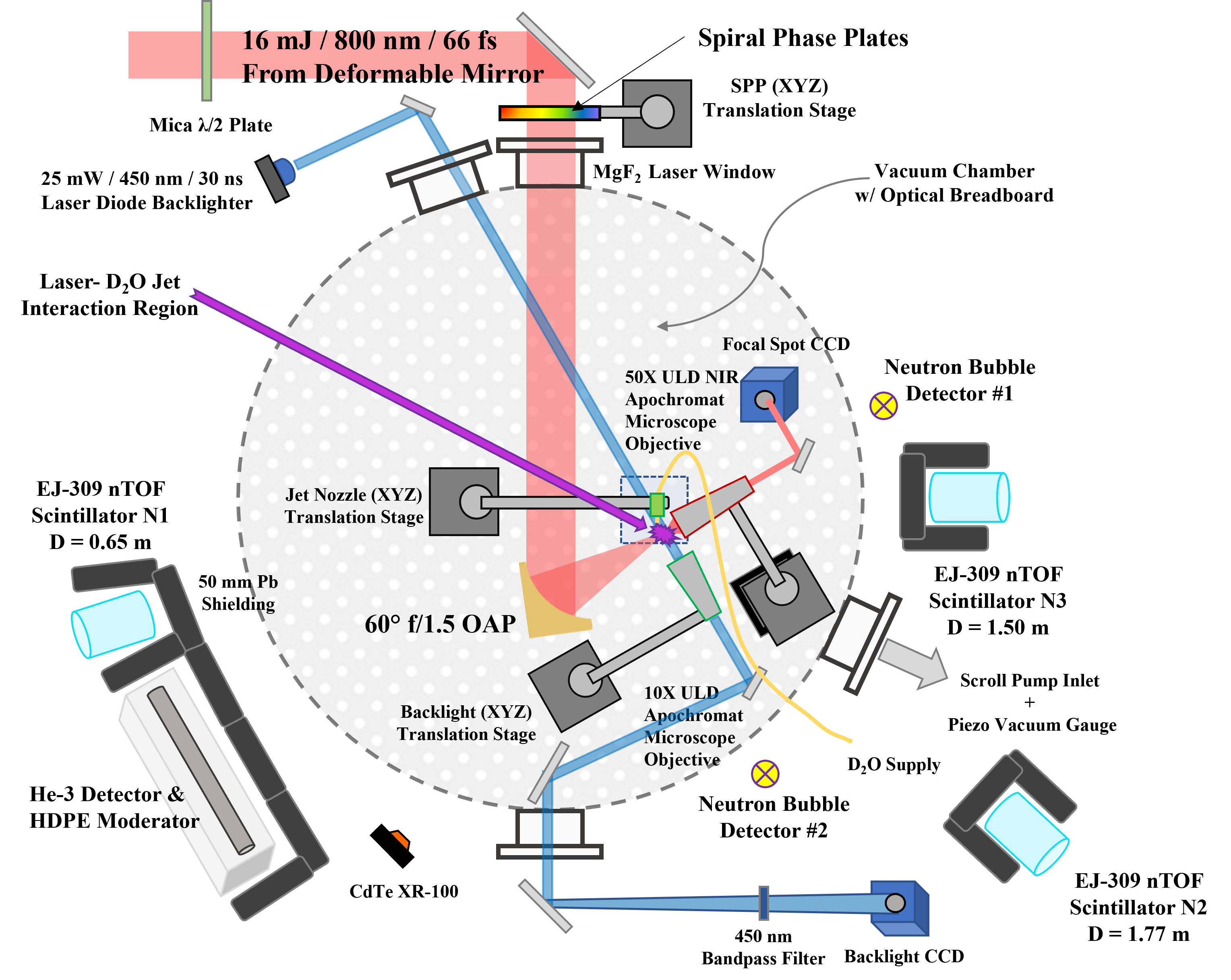}
\caption{The plasma target chamber configuration for experimental investigation of OAM beam effects on ultrashort laser-driven fast neutron yield. The placement of detectors and beam optics is not to scale.} 
\label{1}
\end{figure}

In our experiment, we used two separate spiral phase plates of topological charge $l$ = 1 and 5 to mode convert a normally Gaussian femtosecond laser driver with a center wavelength of $\lambda$ = 800 nm into an OAM beam. Each SPP was fabricated on an optical grade 1.0 mm thick and 47 mm diameter fused silica substrate. The LG$_{01}$ plate was uncoated while the LG$_{05}$ plate was NIR anti-reflection coated. Each plate was retained in a standard lens mount and indexed to the central beam axes. This experiment was performed on the $\lambda^3$ laser system at the University of Michigan. A schematic of the vacuum target chamber, optical diagnostics, and secondary radiation detectors is shown in Figure~\ref{1}. Under experimental conditions, the laser delivered up to 16 mJ of laser energy at $\tau_p$ = 67 fs into a 60$^{\circ}$ f/1.5 Off Axis Parabolic mirror (OAP). This pulse duration was used since the reduced peak power minimizes the risk of catastrophic beam self-focusing in the spiral phase plates inserted in the compressed beam to generate OAM modes. An acoustoptic programmable dispersive filter was also used to pre-compensate for pulse dispersion due to phase accumulation from the driver relay optics and SPP.

Following iterative beam optimization \cite{GAPaper}, the laser was focused to a 2.86 \textmu m 1/$e^2$ Gaussian spot which resulted in a maximum on-target intensity of 7.5 $\times$ 10$^{18}$ W/cm$^2$. All beams were focused on the target as linearly p-polarized light, except as noted in the polarization study where the effects of s-polarized light on the plasma dynamics were investigated by inserting a 100 mm diameter 78 \textmu m thick mica half-wave plate into the driver beam. Similar to work performed by Hah et al. in Ref.~\cite{Hah16} a free-flowing liquid jet was used as the high-repetition rate laser-plasma target. 

To create a stable, laminar jet with a diameter of 20 \textmu m, deuterium oxide of 99.8$\%$ isotopic purity (Sigma-Aldrich, CAS:7789-20-0) was pressurized with a programmable constant displacement/constant pressure syringe pump (Teledyne Iso). To ensure clog-free jet operation, prior to filling the syringe pump reservoir all of the target fluid was passed through 5-micron nylon Luer lock syringe filters. Liquid exiting the pump at 185 \textmu L/min was routed through 300/150 \textmu m O.D./I.D. PEEK tubing and filtered by an in-line 10 \textmu m frit filter. The jet nozzle consisted of a precision cleaved fused silica capillary with 360/15 \textmu m O.D./I.D. (Molex Polymicro) retained in a PEEK CapTite fitting (LabSmith). The microfluidic jet assembly was rigidly held in a 3D-printed mount capable of coupling to standard optomechanical components. Absolute target alignment was performed by manipulating the transverse and longitudinal position of the jet with vacuum compatible closed-loop stepper motors (Thorlabs Z812V). The jet height was set to $\sim$300 \textmu m above the geometric focus of the beam and ensured focused laser pulses were interacting with the laminar cylindrical profile of the fluid jet, well above the position where Rayleigh-Taylor instability would cause stream breakup.
During experiments, the chamber was evacuated with a dry scroll pump to below 12 Torr as recorded on a digital piezo vacuum gauge (Thermovac TM101). D$_2$O was allowed to free-flow into the catch reservoir centered below the jet for a period of 20-30 minutes. A pulsed laser diode (LD) optical backlighting system was used for probing laser-target interactions in this experiment. To further enhance the edge contrast between plasma self-emission and the liquid microjet structure, the central wavelength of the InGaN LD and matched visible filters were changed to $\lambda$ = 450 nm.

Since jet operation was possible at ambient atmospheric conditions, the liquid stream was initialized by starting the microfluidic pump in constant pressure mode (3900-4500 psi) and monitoring the fluid flow rate until the target volume flow rate of 180-190 \textmu L/min was obtained. The pump was then reprogrammed to run in a constant volume flow regime with PID control holding the target flow rate at the empirically optimized value of 185 \textmu L/min. While running at 480 Hz pulse repetition rate, the focus-optimized USPL driver was attenuated to 1-2 mJ and allowed to strike the jet. To obtain optimal placement of the jet position relative to peak focused beam waist, the jet was subsequently translated to maximize hard x-ray dose registered by a calibrated Geiger-Muller detector (Ludlum Model 3003 coupled with a Model 44-7 detector).

\begin{figure}[ht]
  \centering
   \includegraphics[width=8cm]{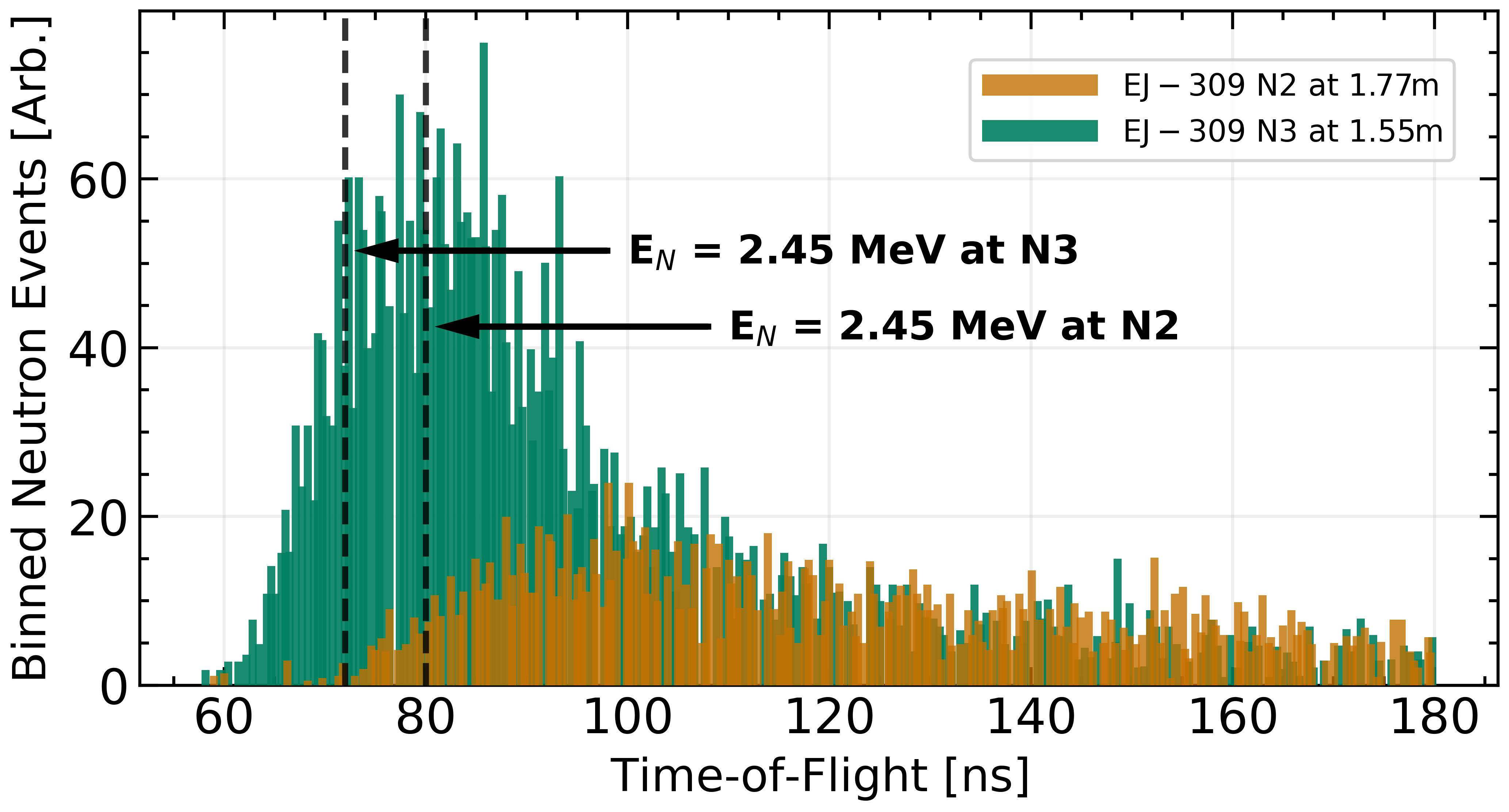}
\caption{Fast neutron time-of-flight measurements performed with $\tau_{coinc}$ = 512 ns coincidence. Liquid scintillation events with proper fast neutron PSP and light output characteristics are shown in 500 ps bins for detectors N2 and N3. Nearly identical signals are recorded for the Gaussian case with $\sim$30\% the number of observed events.  }
\label{10}
\end{figure}

Previous work~\cite{Hah16} observed up to 2.2 $\times$ 10$^5$ n/s/sr D-D fusion neutron yield from a relativistic Gaussian beam interaction with a free-flowing D$_2$O liquid microjet at near normal incidence. An initial attempt to recreate the pre-plasma enhancement noted in this experiment led to confirmation that a grazing incidence laser-target interaction with p-polarized light ultimately corresponds to the highest isotropic neutron flux due to optimized resonance absorption. Here, we investigate grazing incidence interaction of sub-relativistic optical vortices with over-dense free-flowing liquid jet targets at low pressure.
The isotropic D-D neutron flux from grazing incidence interaction with the D$_2$O jet was characterized using the absolute dosimetry provided by neutron bubble detectors and relative counts measured by three calibrated EJ-309 proton-recoil liquid scintillation detectors. During each run, bubble detectors (Bubble Technology Industries BD-PND) were placed at a distance of $d$ = 30.5 $\pm$ 0.5 cm and $d$ = 34.2 $\pm$ 0.5 cm from the laser-plasma interaction region. The scintillation detectors N1-N3, operating in a pulse shape discrimination mode, were located at distances of $d$ = 65.0 $\pm$ 0.5 cm, 177.0 $\pm$ 0.5 cm, and 150.0 $\pm$ 0.5 cm and relative angles to the driver beam of 180$^{\circ}$, 90$^{\circ}$, and 355$^{\circ}$ respectively. The accumulation time for a series of pure Gaussian (PG) and LG$_{05}$ shots at full laser power (16 mJ / 67 fs / 480 Hz) was 300 seconds.
During our experiments, the average isotropic flux measured for beams of topological charge $l$ = 5 over 2.63 $\times$ 10$^6$ laser shots was 8.43(0.17) $\times$ 10$^5$ n/s/sr and the maximum recorded dose during any single run was logged at 1.45(0.34) $\times$ 10$^6$ n/s/sr. In comparison, the average isotropic flux measured for Gaussian beams over 7.78 $\times$ 10$^5$ laser shots was 2.56(0.55) $\times$ 10$^5$ n/s/sr. This lead to a relative signal enhancement of 3.29$\pm$0.71$\times$  for the OAM to PG case as measured by the bubble detectors. Post-processing of the parsed scintillation events corroborated this relative fast neutron signal enhancement and was additionally calculated for each detector as N1: 3.29 $\pm$ 0.69, N2: 3.18 $\pm$ 0.67, and N3: 3.41 $\pm$ 0.72. Figure~\ref{10} provides a representative visualization of the fast neutron signal enhancement seen with high-order OAM shots.

Aside from the difference in magnitude for the total binned events between the LG$_{05}$ and PG case, the nTOF measurement elucidates no fundamental difference in the ion population dynamics. As seen in Figure~\ref{10}, the maximum ion energy cutoff was $\sim$1 MeV and the bulk of the characteristic E$_N$ = 2.45 MeV D-D neutrons arrived at the detector plane with down-scattered energy. These results are consistent with similar laser driven neutron source experiments performed in the presence of a low-pressure background of deuterated vapor~\cite{Hah16,Karsch03}. A lack of extension to the high energy tail of the LG$_{05}$ measured fast neutron spectra suggests the bulk ion temperatures vary little if any from the Gaussian case. This suggests improved volumetric heating likely plays a role in the neutron flux enhancement and validation of this hypothesis was carried out in a computational model.
To further explore the dynamics of OAM enhancement of D-D fusion yield from the free-flowing D$_2$O liquid jet target, a series of transverse scans were conducted. The purpose of this experiment simultaneously investigated the effect three independent variables may play on the measured isotropic fast neutron yield. First, we considered the case of higher peak intensity in the OAM beam mode by introducing an SPP with a topological charge of $l$ = 1.
Second, as is often done in solid-target absorption and x-ray yield experiments~\cite{Hou08,Rajeev02}, we inserted a half-wave plate to rotate the incoming linear polarized driver beam to the s-polarized state. Third, using the LG$_{05}$ optimal longitudinal positions from the previous experiment, we then scanned the spatial position of the target across the tightly focused beams in an attempt to quantitatively characterize what role grazing incidence irradiation plays in the intense vortex beam interaction with the overdense target.
The experimental setup was the same as in the grazing incidence experiment. During each scan, a series of 4.00 \textmu m closed-loop steps were made starting from the farthest transverse position of the $l$ = 5 OAM beam that coincided with the x-ray yield measured above the laboratory background.

At each position, total PSD filtered counts within the fast neutron region of interest from the N1 EJ-309 detector located at 180$^{\circ}$ and $d$ = 65.0 $\pm$ 0.5 cm from the plasma target was used as the detection figure-of-merit. For a dwell time of $t$ = 40.0 seconds, coincidence neutron events from laser-jet interaction were recorded. The chamber pressure was maintained at 12.0 $\pm$ 1.0 Torr, as read out on the digital vacuum gauge, throughout the entire scan. 

\begin{figure}[t!]
  \centering
  \includegraphics[width=8.0cm]{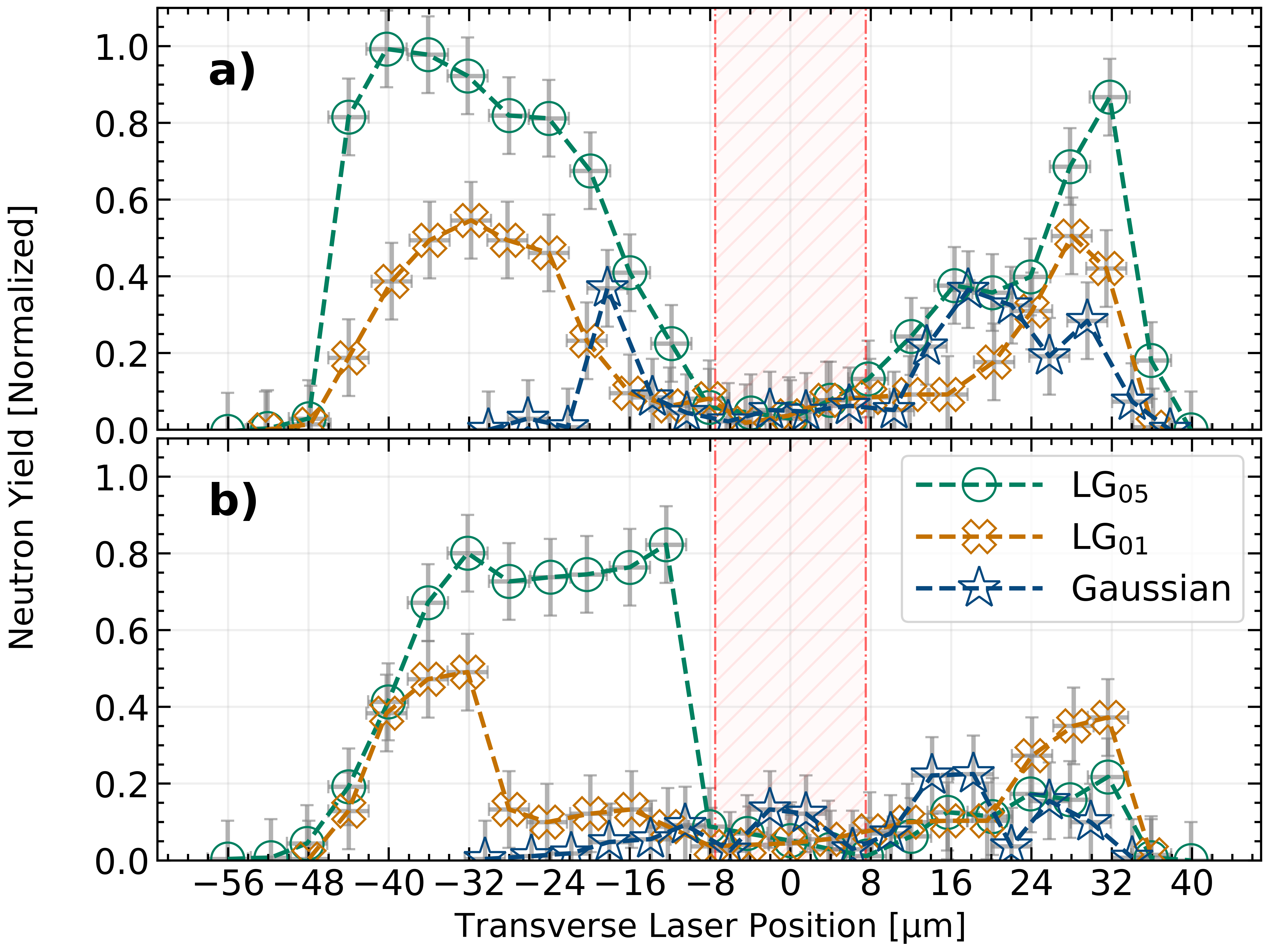}
\caption{Neutron signal normalized to the maximum count recorded for the case of p-polarized light with topological charge $l$ = 5. (a) and (b) plots correspond with the p-polarization and s-polarization of the driver beam, respectively. Hatched red area denotes the approximate bulk water jet position. }
\label{13}
\end{figure}

From Figure~\ref{13}, it is clear that the LG$_{05}$ beam interacts over a spatial scale that extends over $\sim$20 beam diameters (D$_\mathrm{{LG_{05}}}$ = 5.0 \textmu m) and approximately $\sim$7 liquid jet diameters. A similar interaction scale length is also evident for the LG$_{01}$ beam even though the measured intensity profile was approximately 0.44 the diameter of the LG$_{05}$ mode.
Asymmetry shown in the amplitude for the measured relative neutron yield at either edge of the jet for the OAM modes is possibly due to distortion of the annular mode profile (as observed in focal spot imaging) or the incident chirality. All three beams show a pronounced reduction in neutron yield at target-normal incidence independent of the incident beam polarization state. This is expected due to plasma mirroring effects as the target is $n/n_{crit}$ $\approx$ 190. All three scans performed with p-polarized beams show a bi-modal distribution of peak counts across the transverse scan profile but only the spatial rise and fall distances of the PG case are of the order of the 1/$e^2$ beam diameter. The broad spatial range over which the OAM beams efficiently couple energy into plasma surrounding the liquid jet suggests that there is an extended region of critical density ($n_{crit}$ = $m_e\omega^2$/$4\pi e$ = 1.74 $\times$ 10$^{21}$ cm$^{-3}$ for Ti:Sapphire light at $\lambda$ = 800 nm) vapor at the microjet edge. This is unsurprising, as the local density changes five decades in magnitude from approximately $n_e$ = 3.8 $\times$ 10$^{18}$ cm$^{-3}$ to $n_e$ = 3.3 $\times$ 10$^{23}$ cm$^{-3}$ in this same region. 

Following normalization of the binned fast neutron scintillation events to the maximum count measured for the LG$_{05}$ scan, we construct the plot shown in Figure~\ref{13}. In the PG case, the p-polarized light effectively couples into the target through resonance absorption~\cite{Denisov57} at oblique incidence but s-polarized light causes negligible heating across the entire transverse profile. In sharp contrast, efficient laser-plasma energy coupling of an OAM beam having s-polarization is observed for both mode orders and a likely explanation arises from the advancing helical phase which drives skewed propagation of the Poynting vector~\cite{Qiu21}.

PIC simulations are especially important in modeling the computational complexity of relativistic laser physics by means of representative macroparticles, without which the kinetic modeling of liquid jet ion densities $n_i\sim$10$^{23}$ cm$^{-3}$ would be computationally intractable on even the most powerful high-performance computing clusters. Here we utilize the fully relativistic 3-Dimensional electromagnetic PIC code OSIRIS 4.0~\cite{Fonseca02}. 

We modeled the experimental parameters utilized in the $\lambda^3$ shot campaigns albeit with necessary computational truncation of the liquid densities in 3D. We use a spatial interaction volume of 76.4 $\times$ 40.2 $\times$ 40.2 (\textmu m) for the simulation with dimensions of 908 $\times$ 508 $\times$ 508 (X1, X2, X3) cells, correlating to a grid resolution of $\lambda$/10 in each dimension. The time step was set to the CFL condition and demonstrated negligible grid heating with cubic particle weighting for the unperturbed plasma column through the duration of interest. The laser was injected into the simulation from the wall with propagation in the X1 direction. 

The free-flowing D$_2$O jet was initialized as a cylindrical uniform cold plasma with a diameter $\mathrm{D}$ = 15 \textmu m. Jet density was modeled as $n_0$ = 10 $\times$ $n_{crit}$ to relax necessary Debye length resolution conditions and minimize numerical artifacts. The effect of finite laser pre-pulse contrast ($\le$10$^{-7}$) in tandem with a D$_2$O vapor pressure of 12 Torr in the physical experiment induces a pre-plasma and associated density scale length at the liquid jet’s boundary. In order to model the plasma density inhomogeneity, a 3 \textmu m scale length exponential roll off was applied to the particle boundary of the deuterated jet model. Macro particles representing deuterium and oxygen ions were initialized as stationary particles on the grid and assumed to be fully pre-ionized with electrons at 18 particles-per-cell.

\begin{figure*}[htbp]
  \centering
  \includegraphics[width = 450pt]{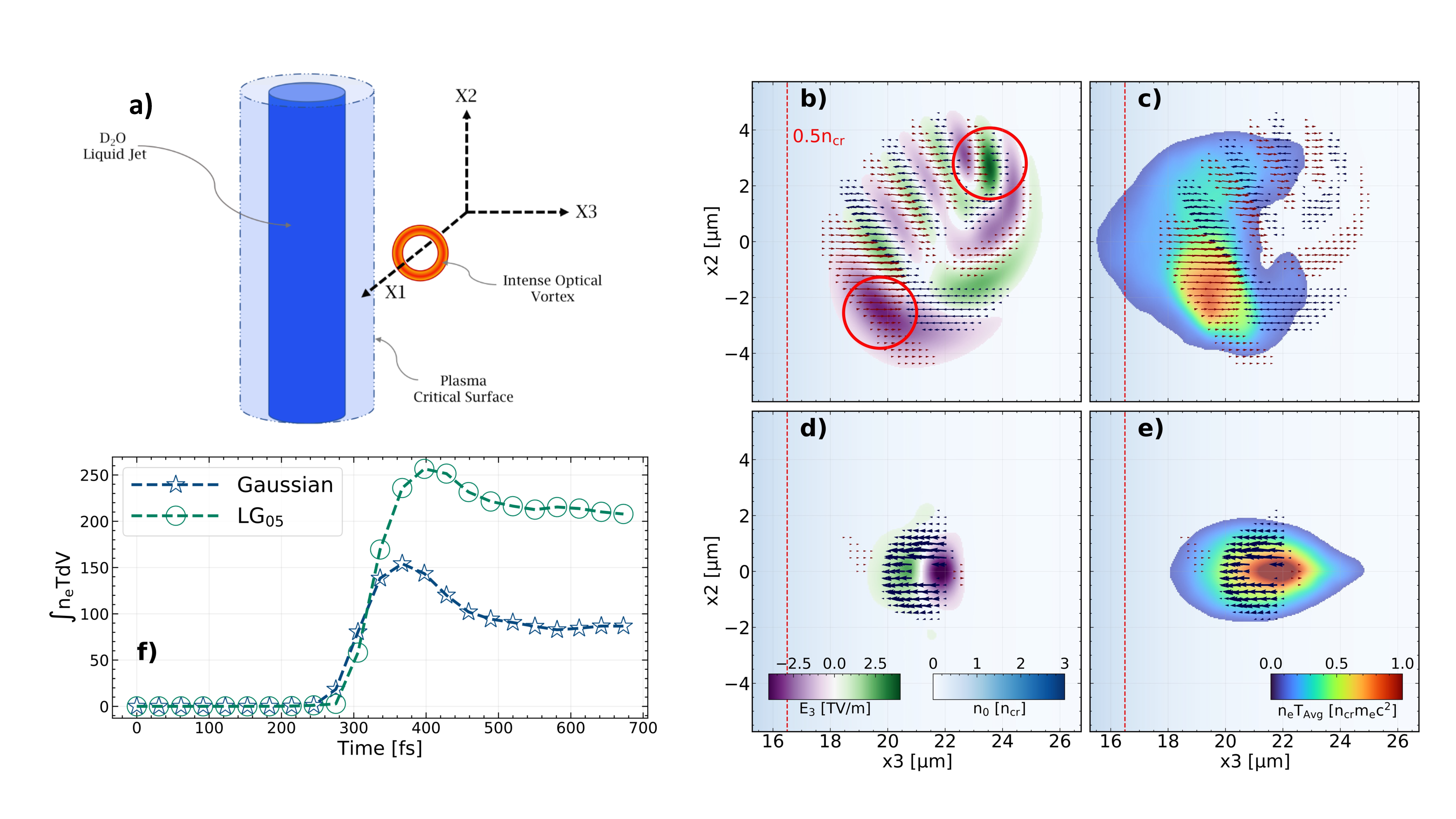}
\caption{(a) A graphical representation of the beam propagation axes and orientation for data slices from the PIC OSIRIS simulation. (b-e) an X2-X3 transverse slice of the LG$_{05}$ and PG beam at offset $r_0$ = 20.0 \textmu m from the jet center. (b,d) demonstrate the p-polarized electric field (green-purple) components with red circles in (b) emphasizing the 2 primary filamentation nodes of the LG beam. Vector arrows demonstrate the X2-X3 currents scaled to their relative magnitudes. Arrows are colored red (blue) for outward (inward) current flows. Blue background depicts the plasma density with a dashed line drawn at the $0.5n_{cr}$ for reference. (c,e) heatmaps show the average electron temperature (averaged across X1), measured in the Eckard rest frame, multiplied by the plasma density with overlaid current arrows. (f) Shows the total volumetric heating of the plasma column for both the PG and LG case over the duration of the simulation.  }
\label{1517}
\end{figure*}

In the simulation, a PG laser pulse with energy, $E_p$ = 16 mJ, $\tau_p$ = 67 fs (FWHM), and a focused beam waist of 2$\omega_0$ = 3.0 \textmu m were used as a relativistic ($a_0$ = 1.5) reference case. Using a built-in suite for OAM beam generation, an LG$_{05}$ beam was launched with equivalent pulse energy, temporal duration, and a properly scaled focused beam waist of $\sim$ 5.5 \textmu m peak-to-peak. As modeled, and in good agreement with theory, the tightly focused LG$_{05}$ mode corresponded with an initial normalized peak vacuum intensity of $a_0$ = 0.67. All laser fields were launched as linear p-polarized light unless otherwise noted. Figure~\ref{1517}(a) shows a basic diagram of the PIC model. Figures~\ref{1517}(b-e) show a comparison of transverse X2-X3 slices taken at $t$ = 92 fs of the PG and LG$_{05}$ beam after propagation at identical radial offsets in the density down-ramp of the liquid jet edge. The slices correspond to the central plane of the water jet taken at the middle of the 67 fs pulse propagation. At this offset distance, the OAM beam has begun self-focusing and self steepening due to the density gradient, leading to the onset of beam filamentation at the nodes of LG profile as previously characterized~\cite{Wilson23,Vuong06,Kant21}. As expected, the Gaussian beam creates an axial channel of hot electrons which ponderomotively expels suprathermal electrons orthogonally into the jet and outward into the low-density background. In contrast, the high-order OAM beam begins to undergo collapse but is limited in its extent and scrapes along the effective critical surface (due to non-normal incidence) in the appearance of an elliptical line-focus with transversely periodic field structures. This flattening along the critical surface creates a large volumetric region of moderate temperature electrons and large amplitude perturbations to the electric field that likely correspond to intense beamlet formation. Under the effects of this spatially optimized self-focusing (which is anticipated to play an even larger role in the presence of an elongated density ramp), the LG$_{05}$ attains relativistic intensity of $a_0$ $\ge$ 1.1 while the PG case exceeds $a_0$ $\ge$ 2.3 . We believe this configuration qualitatively corroborates with significantly enhanced resonance absorption at the critical surface and increasing neutron yield at transverse beam positions exceeding 1.5 jet diameters. The magnitude and structure of currents formed during the LG$_{05}$ beam interaction with the jet are markedly different from those seen in the PG case. The Gaussian beam radially expels electrons forming a starburst-like current flow. Conversely, the self-focusing OAM beam presents with a dense region of rapidly varying, counter-flowing electron streams that strongly resembles the efficient laser-plasma coupling of optimized resonance absorption~\cite{Denisov57,Wilks97}. Indeed, this extended planar region of resonance absorption would lead to a large area of plasma-wave-generated volumetric heating, and the shearing current flow likely caused by helical phase rotation would impart strong plasma oscillations as seen in vortex-like current structures at other points in the simulation. As seen in Figure 4(f), the OAM field allows for enhanced volumetric heating of the jet at laser intensities far below that of a gaussian beam with similar energy. Further modeling and confirmation of the instability growth rate of such an effect could show excellent agreement with the results from Ref.~\cite{Valeo75} where rippling of the critical surface was shown to enhance coupling of non-normally incident laser radiation; but is beyond the scope of the work presented here. This filament-induced rippling overrides the role of the incident polarization state in light-plasma coupling~\cite{Rajeev02} and lends to explain why efficient neutron generation was seen in our experimental campaign for the case of both incident s- and p- polarized beams.  

In conclusion, we demonstrate the application of self-focusing near relativistic optical vortex beams to generate a record isotropic fast neutron yield of 1.45 $\times$ 10$^6$ n/s/sr. Beams with a topological charge of $l$ = 5 were shown to deliver up to a 3.3$\times$ enhancement of fast neutron yield over an optimally tight-focused Gaussian beam. This completely unexpected result was achieved with only modest laser energies of 16 mJ and a relaxed compressed pulse duration of $\tau_p$ = 67 fs. These requirements are well suited for emerging ultrashort pulsed lasers trading the narrower gain bandwidth of Yb-doped laser media for capability of sustained multi-kHz operation delivering kilowatt peak powers \cite{Fattahi2014_FST,Pei_2017, Muller_Limpert_2020_kWfiber,Fan_2021}. Tabletop neutron flux of this level is comparable to that characteristic of commercial sealed-tube D-D generators~\cite{IAEA21} but with a FWHM pulse duration of 20 ns. Further repetition rate scaling (1-10 kHz) of this interaction could easily generate of-order $\sim$10$^{7}$, $\sim$10$^{8}$ n/s, and would be well suited for time resolved prompt gamma neutron activation analysis studies~\cite{Gunther22}. The marked reduction in polarization sensitivity to laser-plasma energy coupling for OAM-target interactions in our physical experiment was investigated through a PIC model and a link to electron density rippling resulting in significant enhancement of plasma wave excitation and laser absorption.

\begin{acknowledgments}
This work was completed with the support of the Department of Energy (DOE) DOE award number DE-NA0002723.  The authors would like to acknowledge the OSIRIS Consortium, consisting of UCLA and IST (Lisbon, Portugal) for providing access to the OSIRIS 4.0 framework. This research was supported in part through computational resources and services provided by Advanced Research Computing at the University of Michigan, Ann Arbor.
\end{acknowledgments}

\bibliography{Peskosky_OAM_2024_bib}

\end{document}